\documentclass[aps,prl,twocolumn,showpacs,superscriptaddress,groupedaddress]{revtex4}  % for review and submission
\usepackage{graphicx}  % needed for figures
\usepackage{dcolumn}   % needed for some tables
\usepackage{bm}        % for math
\usepackage{amssymb}   % for math

\begin{document}

% The following information is for internal review, please remove them for submission
%\widetext
%\leftline{Version xx as of \today}
%\leftline{Primary authors: Joe E. Physics}
%\leftline{To be submitted to (PRL, PRD-RC, PRD, PLB; choose one.)}
%\leftline{Comment to {\tt d0-run2eb-nnn@fnal.gov} by xxx, yyy}
%\centerline{\em D\O\ INTERNAL DOCUMENT -- NOT FOR PUBLIC DISTRIBUTION}

% the following line is for submission, including submission to the arXiv!!
%\hspace{5.2in} \mbox{Fermilab-Pub-04/xxx-E}

\title{Detonative Propagation and Accelerative Expansion of the
Crab Nebula Shock Front}
%\input author_list.tex       % D0 authors (remove the first 3 lines
                             % of this file prior to submission, they
                             % contain a time stamp for the authorlist)
                             % (includes institutions and visitors)
\affiliation{Center for Combustion Energy and Department of Thermal Engineering, \\Tsinghua University, Beijing 100084, China}
\affiliation{Department of Mechanical and Aerospace Engineering, Princeton University, Princeton, NJ 08544-5263, USA}
\author{Yang Gao$^1$}
%  \affiliation{Center for Combustion Energy and Department of Thermal Engineering, Tsinghua University, Beijing 100084, China}
\author{Chung K. Law$^{1,2,\ast}$}
% \affiliation{Center for Combustion Energy and Department of Thermal Engineering, Tsinghua University, Beijing 100084, China}
% \affiliation{Department of Mechanical and Aerospace Engineering, Princeton University, Princeton, NJ 08544-5263, USA}

\date{\today}

\begin{abstract}
  The accelerative expansion of the Crab nebula's outer envelope is a mystery in dynamics as a conventional expanding blast wave decelerates when bumping into the surrounding interstellar medium. Here we show that the strong relativistic pulsar wind bumping into its surrounding nebula induces energy-generating processes and initiates a detonation wave that propagates outward to form the current outer edge, namely the shock front, of the nebula. The resulting detonation wave, with a reactive downstream, then provides the needed power to maintain propagation of the shock front. Furthermore, relaxation of the curvature-induced reduction of the propagation velocity from the initial state of formation to the asymptotic, planar state of Chapman-Jouguet propagation explains the observed accelerative expansion. The essential role of detonative propagation in the structure and dynamics of the Crab nebula offers potential richness in incorporating reactive fronts in the description of various astronomical phenomena.
\end{abstract}

\pacs{95.30.Lz, 98.58.Mj, 47.40.Rs}
% astro: hydrodynamics, astro: supernova remnant, fluid dynamics: detonation
\maketitle

%\section{\label{sec:level1}First-level heading}
% sections are not used for PRL papers

%\paragraph*{Introduction}
The Crab nebula is the most observed supernova remnant since its birth in A.D. 1054. Its explicit birth date, together with accurate observations of its subsequent evolvement through optical \cite{duncan1939,trimble1968,nugent1998} and radio \cite{bietenholz1991} facilities in the last several decades, has yielded rich information on its structure and dynamics. In particular, it is shown that this nebula has two major features, namely optical line-emitting filaments, which are supposed to be the dense condensations of the thermal gas \cite{trimble1968,nugent1998}, and an expanding radio synchrotron bubble shell, which is recognized as the outer edge of the nebula composed of light relativistic gas and magnetic fields \cite{bietenholz1991}. Furthermore, using observations from 1939 to 1992, and by assuming a constant expansion speed for the nebula, the date of the supernova outburst was estimated  to be A.D. 1130$\pm$16 \cite{nugent1998}, which is 76 years after the historically observed date of A.D. 1054. This discrepancy then implies that the nebula expansion is actually accelerative. A similar result was obtained through an analysis of the nebula's radio shell \cite{bietenholz1991}, at $r_{\rm d}\approx2~{\rm pc}=6\times10^{18}~{\rm cm}$. Based on the most recent result that the current expansion velocity of the nebula shell is $v_{\rm p}\sim 1.5\times10^8~{\rm cm~s^{-1}}$, and if a constant acceleration $\dot{v}_{\rm c}$  is assumed, one can then estimate that the acceleration is $0.82\times10^{-3}~{\rm cm~s^{-2}}$.
%\textbf{We may infer from these quantities that the acceleration of the Crab nebula is low, by only $\sim$ 5\% for its whole life of %$\sim$ 1,000 years.}

	The above observation, however, cannot be satisfactorily explained by current understanding of the dynamics of the expanding front, on two counts, both based on the energy source needed to sustain the front propagation. First, it has been suggested that the energy from the spin down of the central pulsar provides the needed energy source \cite{bejger2003,manchester1977}. However, it is not clear as how the spin-down energy of the central pulsar, carried by the pulsar wind, can be transferred to the kinetic energy needed to support the expansion of the nebula's outer edge. This is because the matching radius of the strong MHD shock, $r_{\rm s}\approx 3\times10^{17}~{\rm cm}=0.1~{\rm pc}$, formed by the highly relativistic pulsar wind bumping into the surrounding nebula, is much smaller than the outer edge of the nebula, $r_{\rm d} \approx 2~{\rm pc}$, to induce effective coupling (see Fig. 1 for an illustration of these radii). Second, present theoretical considerations generally describe the outer edge of the Crab nebula as the front of a (magneto) hydrodynamic blast wave that bumps into the surrounding interstellar medium (ISM) \cite{kennel1984}. However, since a conventional expanding blast wave decelerates \cite{landau1959} as it transfers kinetic energy to its surrounding, such a feature not only contradicts the accelerative nature of the nebula's outer shell, it cannot even support the steady expansion of the front.
It is emphasized that the acceleration, while low in magnitude, being just $\sim5\%$ in the entire life of $\sim$1,000 years, is nevertheless finite and as such accentuates the observed absence of deceleration.

In view of the above concerns, it is of intrinsic interest to explore whether there are other energy sources and coupling mechanisms for the observed accelerative expansion of the nebula shock front. We suggest herein that the shock front is actually a detonation wave, which is a shock wave with an exothermic, reactive downstream. The exothermic heat release then provides the needed energy for the sustenance of the propagation. Furthermore, relaxation of the front curvature upon expansion facilitates the front propagation, thereby resulting in the observed accelerative expansion. We now present our interpretation and implementation of the detonative description of the nebula shock front propagation, with corresponding estimates of the various dynamic and thermodynamic parameters supporting such a mechanism.

\begin{figure}
\includegraphics[scale=0.33]{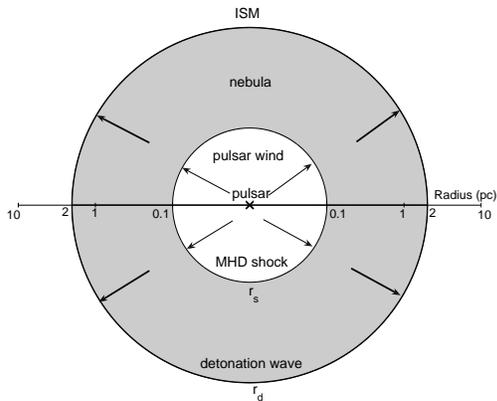}
\caption{Schematic structure of the Crab nebula. The cross in the center represents the pulsar. The region immediately around the pulsar is the pulsar wind domain, bounded by the surrounding nebula with an interface of strong MHD shock at $r_{\rm s} = 0.1~ {\rm pc}$. Beyond this radius is the nebula region, with an outer edge at $r_{\rm d} = 2~ {\rm pc}$, where the detonation wave forms as the nebula expands into the ISM.}
\end{figure}

\paragraph*{Detonation wave at the nebula envelope}
To facilitate appreciation of the role of the detonation wave on the problem at hand, it is instructive to first discuss its structure, initiation and propagation based on understanding largely gained from studies of reactive gasdynamics \cite{landau1959,law2006}.  Briefly, in the wave stationary frame, a detonation wave consists of a leading hydrodynamic shock which slows the incoming supersonic flow to subsonic and changes its thermodynamic states according to the equation of state of the medium. This leads to the Neumann state immediately downstream of the shock. For conventional materials including that of interest in the nebula, the density, pressure and temperature are increased upon crossing the shock, which induces reaction of the material after an ignition delay length, $l_{\rm ig}$. This forms the ZND structure of the detonation wave. For reactions that are highly temperature sensitive, the subsequent emergence of the reactions is abrupt and basically complete, in the manner of a step function. Figure 2 shows a schematic of the profiles of the various properties.

	Propagation of a detonation wave is affected by the downstream rarefaction waves that continuously overtake the shock front and weaken its strength until the velocity downstream of $l_{\rm ig}$ becomes sonic. For 1D planar propagation without loss, the sonic state is attained at downstream infinity and steady-state propagation is attained, resulting in the Chapman-Jouguet (C-J) mode of propagation. This feature of detonation propagation therefor assures the non-deceleration of the Crab nebula shock front. The acceleration, however, is due to the relaxation of the so-called curvature effect. Specifically for an expanding front the divergence of the streamline downstream of the shock slows down the flow velocity, advances the sonic state from far downstream towards the front, and consequently reduces the propagation velocity from the C-J value, with the reduction increasing with increasing front curvature. This curvature-induced reduction has three implications. First, there exists a minimum, critical radius $R_{\rm cr}$ for the initiation of the detonation propagation. Second, once initiated and the front curvature relaxes as it expands, the propagation velocity would continuously \emph{increase} to approach the C-J value. Third, a minimum, critical energy is needed for a detonation to attain its critical radius.

We shall now apply the above understanding to analyze the propagation of the nebula front.   Referring to Fig. 1, it is proposed that from the beginning of the nebula's evolution, the pulsar wind has been pushing its surrounding medium outward continuously until the interface expanded to $r_{\rm s}$, at which a forward and a reverse detonation fronts are simultaneously formed, with the forward front propagating outward and the reverse front propagating inward. The continuously expanding, forward detonation wave then forms the outer envelope of the Crab nebula observed today. The motion of the latter, however, is rapidly arrested by the outwardly directed pulsar wind, forming the MHD shock. The radius $r_{\rm s}$, at which the reverse detonation front is stabilized, can thus be identified as the critical radius for detonation initiation, which we shall discuss in following context.

Our analysis is based on a spherically symmetric model describing the roughly spherical nebula's outer shell as was observed \cite{bietenholz1991}. As a first attempt to the detonation interpretation of the nebula's expanding shell, we do not consider the filament structure of the nebula region \cite{trimble1968,nugent1998}, but will make some observations of its possible effects later. We further note that previous analysis \cite{kennel1984} shows that in order to convert the kinetic energy flux effectively into the observed synchrotron luminosity, the ratio of the magnetic field energy to the particle kinetic energy should be very small. Then it is reasonable to neglect the magnetic field when considering the energy components in this wind-nebula shock at $r_{\rm s}$, as well as in the more distant nebula regions where the detonation takes place. In addition, we also assume that the gravitational energy is much smaller than the kinetic energy of the flow.

With the above assumptions, the problem simplifies to that of the structure and propagation of a spherically expanding detonation wave which have already been well analyzed \cite{he1994,yao1995}, and was presented in \cite{law2006}. We shall therefore simply state the key results relevant for the present problem. Through estimates of the various dynamic and thermodynamic parameters governing the present process, we shall demonstrate that the observed phenomena associated with the expanding nebula front are consistent with the description based on detonation theory.

\paragraph*{Parametric structure of the detonation wave}
We define the radius $R_{\rm s}$, the propagation speed D, the heat of reaction $q_{\rm c}$ and the reaction induction length $l_{\rm ig}$ for the detonation wave. The corresponding dimensionless variables are $\hat{l}_{\rm ig}=l_{\rm ig}/R_{\rm s}$, $\hat{q}_{\rm c}=q_{\rm c}\gamma/a_{\rm u}^2$ and $\hat{D}=D/a_{\rm u}$ the Mach number of the detonation velocity, where $a=\sqrt{\gamma p/\rho}$ is the sound speed, $\gamma$ is the adiabatic index and the subscript $u$ denotes the upstream ISM. Focusing on the evolution of the detonation Mach number, we can express it as $\hat{D}=\hat{D}_{\rm CJ}(1-\Delta)$, where $\Delta \ll 1$ is a small number representing its deviation from the asymptotic C-J value caused by the curvature stretch. Then we have \cite{law2006}
  \begin{equation}
  \hat{D}_{\rm CJ}^2=2(\gamma+1)+2(\gamma^2-1)\frac{\hat{q}_{\rm
  c}}{\gamma} \approx 2(\gamma^2-1)\frac{\hat{q}_{\rm c}}{\gamma},
  \label{equ:D_cj}
  \end{equation}
  \begin{equation}
  \Delta=\frac{8\gamma^2}{\gamma^2-1}\hat{l}_{\rm ig}.
  \label{equ:delta}
  \end{equation}
for large values of $\hat{q}_{\rm c}$ relevant for the present situation of strong detonation, which has an estimated Mach number of 170 (see Table I). The typical temperature of the unburned upstream ISM is \cite{ferriere2001} $T_{\rm u}\sim 10^4~{\rm K}$ and the density is $\rho_{\rm u} \sim 0.5~{\rm cm^{-3}}$. Other parameters of the upstream ISM are readily calculated and listed in Table I. Furthermore, according to eq. (1), the heat of reaction that supports the expansion of the detonation wave is estimated as
    \begin{equation}
  q_{\rm c}\approx5.9\times10^{12}~{\rm J/mol}=61~{\rm MeV/atom},
  \label{equ:reaction heat}
  \end{equation}
  which is relatively large compared to typical values of light nuclear reactions and those of chemical reactions.

It is also noted that in the above calculations, the adiabatic index of the upstream ISM gas has been assumed to be  $\gamma=1.1$, resulting in a large thermal capacity, $c_{\rm p}\approx c_{\rm v}\approx 10 R$, with $R$ being the molar gas constant. The index $\gamma=1.1$ is intermediate of isothermal ($\gamma=1$) and adiabatic ($\gamma=5/3$ for monatomic gas) fluids, and is a typical value for ISM with partially effective thermal conduction. Furthermore, detonation cannot occur for isothermal ISM gas with $\gamma=1$,  while for the adiabatic limit case of $\gamma = 5/3$, the reaction energy release $q_{\rm c}$ will decrease to $\sim$1/6 its current value, and the velocity deviation $\Delta$ will decrease to 1/4 its current value.
These variations will cause tuning of the detonation parameters, but will not change the principal
  conclusions made in this letter.

We can further calculate the physical parameters describing the Neumann state just downstream of the shock front but before the reactions are initiated \cite{law2006}. After the reactions have taken place, the downstream (burned) gas will eventually approach the C-J state, whose parameters can also be readily determined. Parameters of the upstream state, the Neumann state and the burned C-J state are listed in Table I and schematically shown in Fig. 2.

It is noted that under the Neumann temperature $T_{\rm N}= 1.4\times10^7~{\rm K}$, some nuclear reactions, e.g. the p-p cycle \cite{harwit1998}, are able to be initiated. However, as the density in the nebula region is smaller by a factor of $\sim 10^{-25}$ compared to that in the stellar central regions, reaction rates of these nuclear processes are very small. Since typical heats of chemical reaction are also very small, alternative energy-generating processes need to be identified to supply the energy of $q_{\rm c}\approx61~{\rm MeV/atom}$. One possible energy source for the detonation is the inverse Compton process \cite{gould1965,dejager1992}, which is a commonly used mechanism in astrophysics. Crab nebula's high energy $\gamma$-ray emissions and flares are supposed to be produced by inverse Compton scatterings \cite{abdo2011,tavani2011}, and very probably, part of the high energy photons will interact with the nebula gas to support expansion of the envelope.

\begin{table}
\begin{center}
\caption{Gas parameters of the detonation front: upstream ISM (u), Neumann state (N) and burned C-J state (C-J).}
\begin{tabular}{ccccccc}
\hline
    & $T$    & $u$ & $\rho$ & $p$ & $a$ & $M$ \\
    &(K)   & $({\rm km~s^{-1}})$ & $({\rm cm^{-3}})$ & $({\rm Pa})$ & $({\rm km~s^{-1}})$ & \\
\hline
u   & $10^4$          &1500    &0.5     &$6.2\times10^{-14}$ &9   &170  \\
N   & $1.4\times10^7$ & 74     &10      &$1.6\times10^{-9}$  &340 &0.22 \\
C-J  &$7.5\times10^7$  &790     &1.1     &$0.9\times10^{-9}$  &790 &0.9  \\
\hline
\end{tabular}
%: temperature, velocity, density, pressure, sound speed and Mach number of the upstream ISM, the compressed Neumann gas and the burned %C-J gas.}
\end{center}
\end{table}

\begin{figure}
\includegraphics[scale=0.36]{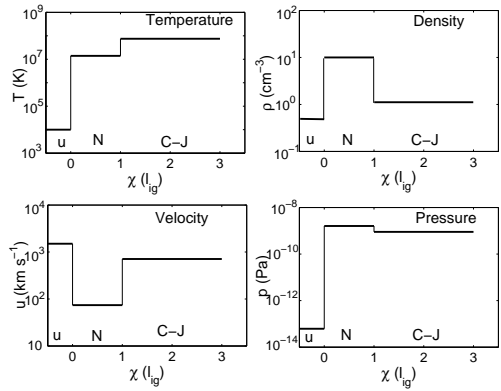}
\caption{Temperature, density, velocity and pressure of the upstream ISM, the compressed Neumann state and the burned C-J gas are shown as step functions in each panels. The Neumann state crosses the induction length of $l_{\rm ig} = 1.0\times10^{-4}~{\rm pc}$, while the upstream and burned gas extends outward to infinite and inward to the inner edge of the nebula, respectively.}
\end{figure}

\paragraph*{Initiation of the detonation wave }
The standing inward shock front (see Fig. 1 and reference \cite{kennel1984}) is assumed to be the radius where the forward and reverse detonations began to form and separated in the early days of the nebula evolution.
%So the critical radius for the formation of the forward propagating detonation can be identified as that of the standing MHD reverse %shock, namely $R_{\rm cr} = r_{\rm s} = 0.1~ {\rm pc}$.
Since the reverse shock is expected to be immediately captured by the pulsar wind, the radius of its front varies little since the formation of the binary detonations.
 %So the above estimate of critical radius for the outgoing detonation is reasonable.}
 Consequently the critical radius for the formation of the forward propagating detonation can be identified as that of the standing MHD reverse shock, namely $R_{\rm cr} = r_{\rm s} = 0.1~ {\rm pc}$.

 Directly referring to eq. (14.8.29) in \cite{law2006} that relates the induction length and the critical radius, and assuming $\beta=4$, which is consistent with the reaction activation temperature estimated based on the Neumann state, the induction length can be estimated as
  \begin{equation}
  l_{\rm ig}\sim R_{\rm cr}\bigg(\beta\frac{16 e
  \gamma^2}{\gamma^2-1}\bigg)^{-1} \approx 1.0\times10^{-4}~{\rm pc},
  \label{equ:induction}
  \end{equation}
which is a small number compared to the radius of the detonation wave, $r_{\rm d} = 2~ {\rm pc}$, and is therefore consistent with the requirement of the formulation. We then have the quantitative expression of eq. (2):
  \begin{equation}
  \Delta=\frac{8\gamma^2}{\gamma^2-1}\hat{l}_{\rm ig}\approx\frac{0.004~{\rm pc}}{R_{\rm s}}.
  \label{equ:delta2}
  \end{equation}

Attainment of the critical radius for the detonation initiation requires the initial blast wave to possess a corresponding critical energy $E_{\rm cr}$ \cite{zeldovich1985}.  Referring to eq. (14.8.30) in \cite{law2006}, we find:
  \begin{equation}
  E_{\rm
  cr}=\bigg(\frac{8je\beta\gamma^2}{\gamma^2-1}\bigg)^{j+1} k_{\rm
  j} \rho_{\rm u} D_{\rm CJ}^2 l_{\rm ig}^{j+1}\approx
  3\times10^{45}~{\rm erg},
  \label{equ:critical e}
  \end{equation}
 where $j= 2$ and $k_{\rm j}=5.31$ for the spherical geometry. This critical energy is much smaller than the part of pulsar spin-down energy carried by the pulsar wind, estimated to be $\sim 10^{49}~{\rm erg}$ \cite{kennel1984,hester2008}. So the energy carried by the relativistic pulsar wind is sufficient to initiate the forward detonation wave. For those supernova remnant systems whose pulsar wind energies are not large enough to reach the critical value of $E_{\rm cr}\sim 3\times10^{45}~{\rm erg}$, detonation \emph{cannot} be successfully initiated; instead, it will decelerate as it defuses into the surrounding medium. We can therefore classify those nebulae around supernovae into two groups: namely acceleratively expanding ones led by detonation waves, and defused ones formed by decelerating blast waves.

\paragraph*{Accelerative expansion of the detonation front}
Subsequent to attaining the critical radius, the forward detonation then propagates with the velocity
  \begin{equation}
  D=D_{\rm CJ}(1-\Delta)=1.5\times 10^3\times
  (1-\frac{0.004~{\rm pc}}{R_{\rm s}})~{\rm km~s^{-1}}
  \label{equ:detonation speed}
  \end{equation}
in the nebula domain that begins at $r_{\rm s} = 0.1~ {\rm pc}$ and ends at the current nebula envelope at $r_{\rm d} = 2~ {\rm pc}$. The variation of the nebula envelope velocity as a function of the envelope radius is shown in Fig. 3, demonstrating an accelerative expansion of the wave front. It is seen that at the early age of the nebula, when $R_{\rm s}$ is slightly larger than 0.1 pc, the propagation velocity of the detonation is smaller than its present value by about 5 percent. Furthermore, the detonation velocity accelerates more rapidly in the early days of the front formation, and gradually approaches the C-J velocity as it evolves. We have therefore successfully explained the accelerative propagation of the nebula front.

\begin{figure}
\includegraphics[scale=0.33]{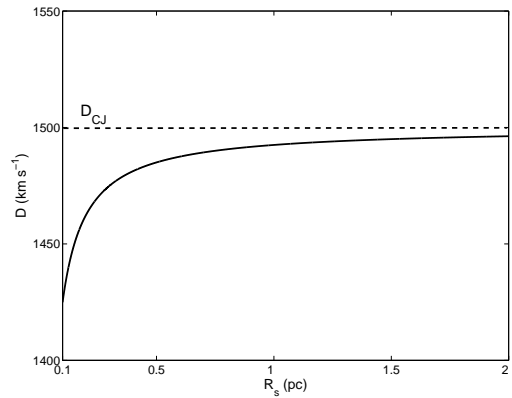}
\caption{Increase of the detonation propagation velocity as a function of the wave front radius. The detonation velocity is $\sim1425~{\rm km~s^{-1}}$ at $R_{\rm s}=0.1~{\rm pc}$ where the detonation begins and gradually approaches
  the C-J velocity of $\sim1500~{\rm km~s^{-1}}$  as the wave expands.}
\end{figure}

\paragraph*{Summary and Discussion}

In this first attempt to integrate detonation features in the structure and dynamics of the Crab nebula, we have demonstrated that the concept is a viable one in terms of the observed phenomena as well as the various estimated dynamic and thermodynamic parametric values associated with the description. In particular, the following understanding has been gained. (1) There is a supersonic detonation wave at the front of the expanding nebula with a Mach number of $\sim$170; the wave is supported by downstream energy-generating reactions. (2) The surrounding ISM is compressed to a temperature of $10^7-10^8~{\rm K}$ and a density of $1-10~{\rm cm^{-3}}$ by the initial shock and the subsequent reactive heating. (3) The envelope of the nebula is accelerating due to relaxation of the front curvature as it evolves to the planar limit upon continuous expansion.

We close this discourse with some additional considerations. First, there is another possible mechanism involving the pulsar wind energy that may account for the accelerative expansion of the Crab nebula outer envelope. That is, the downstream reaction of the inward detonation may last for a long time, in which case the delayed exothermic reaction takes place close to the outer envelope and provides the kinetic energy needed for its accelerative expansion. This mechanism is similar to the flame acceleration in obstructed channels \cite{bychkov2008}. However, it is an open issue to confirm the existence of delayed exothermic reactions in the Crab nebula.

Second, it is noted that the filament structure of the nebula could be related to the well-established cellular morphology of detonation fronts \cite{law2006}. These cells are consequences of interacting triple-shock ensembles, which in turn generate complex local subsonic and supersonic flows that could be manifested as filaments downstream of the front. The presence of cells could also increase the area of the front and consequently its global propagation rate. If the increase of the flame area is itself accelerative, then it could constitute an additional mechanism for the observed accelerative expansion of the nebula.

It is also of interest to investigate whether the detonation structure suggested here exists in other supernova remnants (SNRs), such as those well-observed SNRs in different evolution phases.  Applications of detonation and deflagration theories to describe the chemical or nuclear processes in all kinds of ISM shocks are also merited.

%\begin{figure}
%\caption{Each figure legend should begin with a brief title for
%the whole figure and continue with a short description of each
%panel and the symbols used. For contributions with methods
%sections, legends should not contain any details of methods, or
%exceed 100 words (fewer than 500 words in total for the whole
%paper). In contributions without methods sections, legends should
%be fewer than 300 words (800 words or fewer in total for the whole
%paper).}
%\end{figure}

%\textit{Physical Review} style requires that the initial citation of
%figures or tables be in numerical order in text, so don't cite
%Fig.~\ref{fig:wide} until Fig.~\ref{fig:epsart} has been cited.

%\input acknowledgement.tex   % input acknowledgement

This work was supported by the startup fund for the Center for Combustion Energy at Tsinghua University.

\end{document}